\documentclass[rnote]{aa} 

\usepackage{graphicx}
\usepackage{txfonts}

\newcommand{\lsun}{L$_{\odot}$}
\newcommand{\msun}{M$_{\odot}$}

\newcommand{\feii}{[\ion{Fe}{ii}]}
\newcommand{\sii}{[\ion{S}{ii}]}
\newcommand{\pii}{[\ion{P}{ii}]}
\newcommand{\um}{$\mu$m}

\newcommand{\lbol}{$L_{\mathrm{bol}}$}

\begin{document}

   \title{The HH34 outflow as seen in \feii1.64$\mu$m by LBT-LUCI\thanks{Based on observations obtained with the Large Binocular Telescope.}}

   \author{S. Antoniucci
          \inst{1}
          \and
          A. La Camera\inst{2} \and B. Nisini\inst{1} \and T. Giannini\inst{1}
          \and D. Lorenzetti\inst{1} \and D. Paris\inst{1} \and E. Sani\inst{3}
          }

   \institute{INAF - Osservatorio Astronomico di Roma, Via di Frascati 33, 00040 Monte Porzio Catone, Italy \and
   DIBRIS, Universit\`a di Genova, Via Dodecaneso 35, 16146, Genova, Italy \and
   INAF-Osservatorio Astrofisico di Arcetri, Largo Enrico Fermi 5, 50125 Firenze, Italy}
   \date{Received ; accepted }
   \offprints{Simone Antoniucci, \email{simone.antoniucci@oa-roma.astro.it}}

 
  \titlerunning{HH34 as seen in \feii1.64$\mu$m by LBT-LUCI} 
 
  \abstract
   {Dense atomic jets from young stars copiously emit in \feii\ IR lines, which can, therefore,
   be used to trace the immediate environments of embedded protostars.}
   {We want to investigate the morphology of the bright \feii\ 1.64$\mu$m line 
   in the jet of the source HH34 IRS and compare it with the most
   commonly used optical tracer \sii.}
   {We analyse a 1.64$\mu$m narrow-band filter image obtained with the Large Binocular Telescope (LBT) LUCI instrument,
   which covers the HH34 jet and counterjet. A Point Spread Function (PSF) deconvolution algorithm was applied
   to enhance spatial resolution and make the IR image directly comparable to a \sii\ HST 
   image of the same source. 
    }
   {The \feii\ emission is detected from both the jet, the (weak) counter-jet, and from the HH34-S and HH34-N bow shocks. The deconvolved image allows us to
   resolve jet knots close to about 1\arcsec from the central source. The morphology of the 
   \feii\ emission is remarkably similar to that of the \sii\ emission, and the
   relative positions of \feii\ and \sii\ peaks are shifted according to proper motion measurements,
   which were previously derived from HST images. An analysis of the \feii/\sii\ emission ratio shows that Fe gas abundance is much lower
   than the solar value with up to 90\% of Fe depletion in the inner jet knots. This confirms previous
   findings on dusty jets, where shocks are not efficient enough to remove refractory species from
   grains.}
   {}


   \maketitle
%

   \begin{figure}[!h]
   \centering
\includegraphics[angle=0,width=7.5cm]{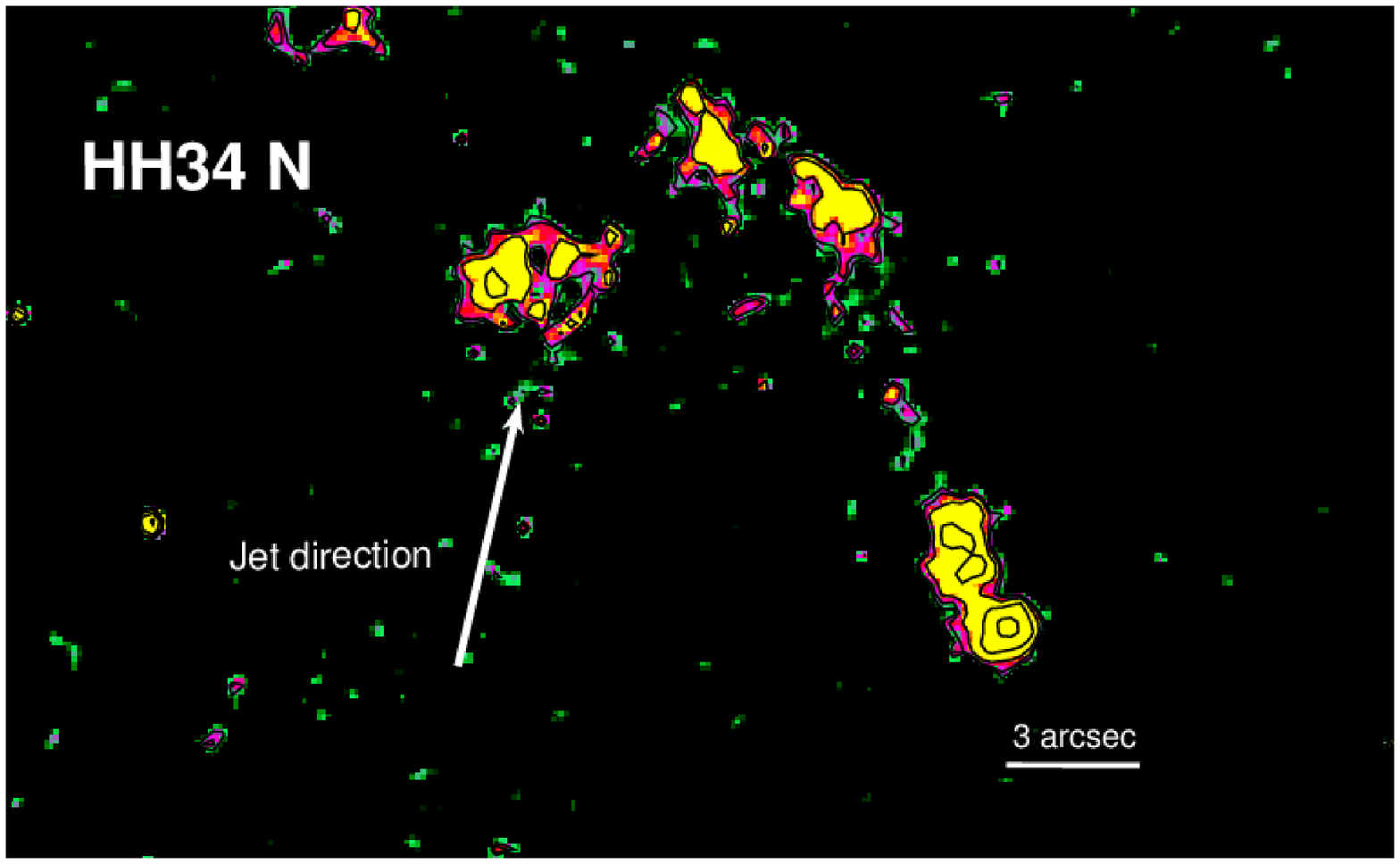}
\includegraphics[angle=0,width=\hsize]{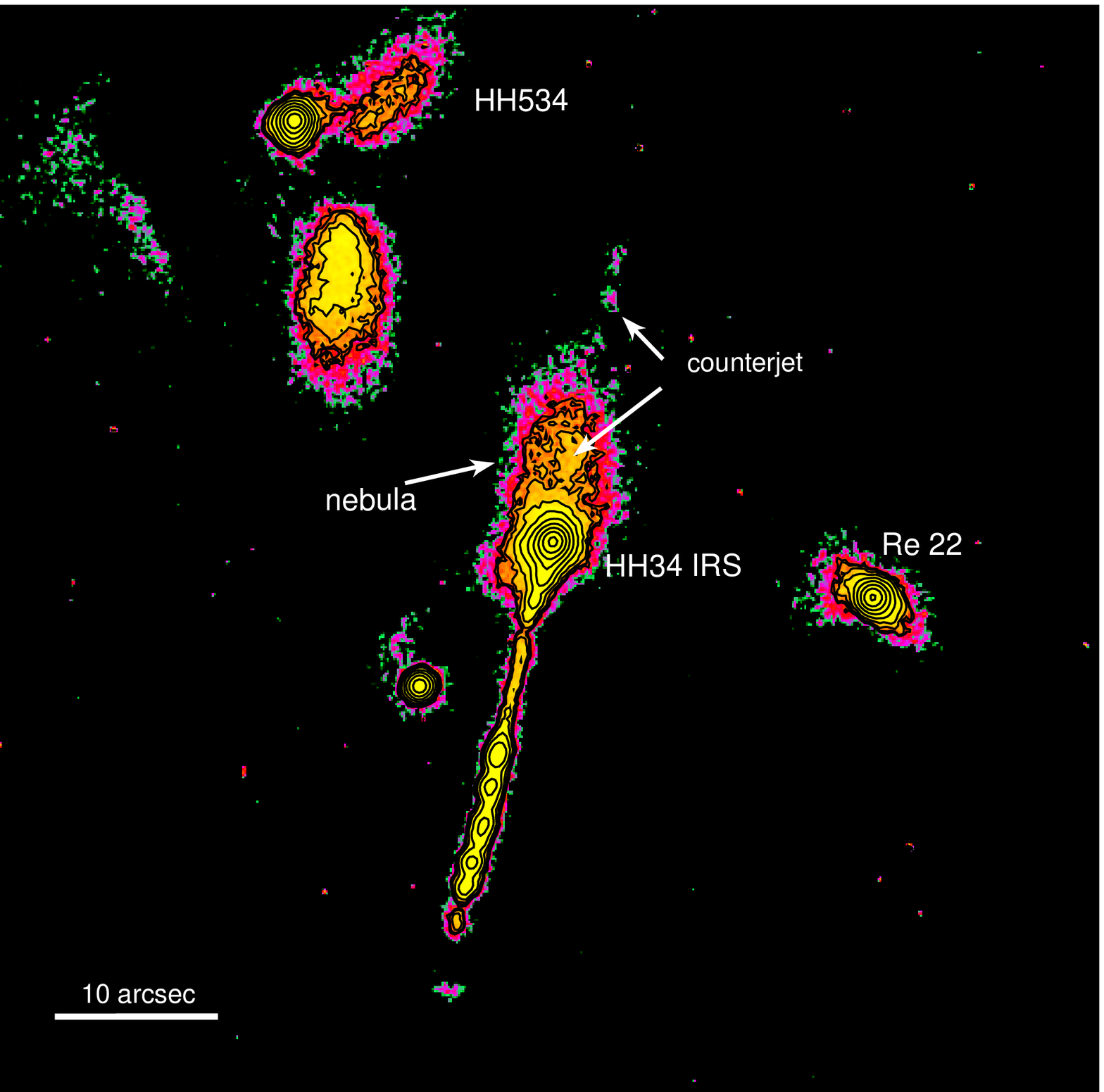}\\
\vspace{0.1cm}
\includegraphics[angle=0,width=7.5cm]{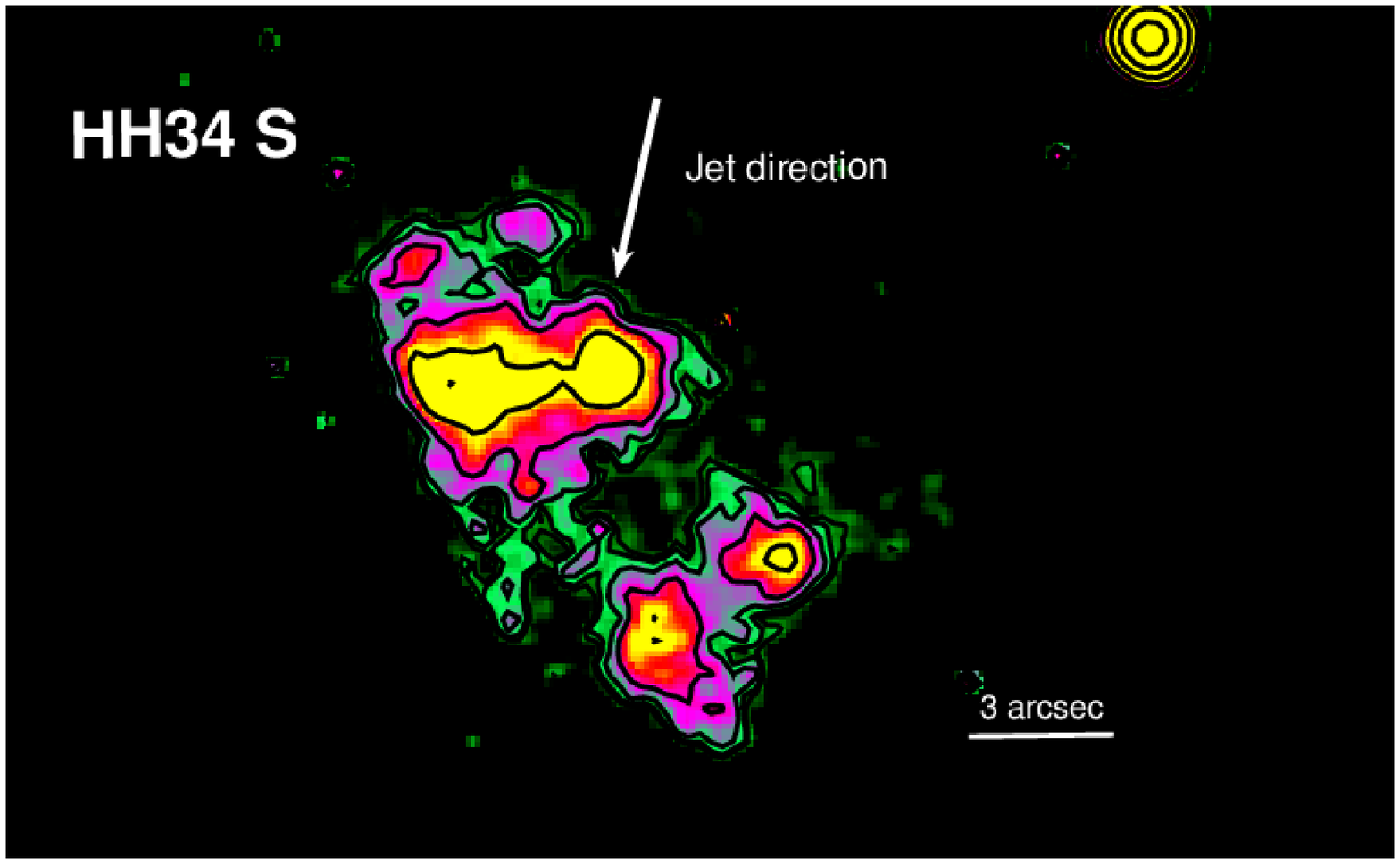}
    \caption{LBT-LUCI image of the HH34 region obtained with a narrow-band filter
    centred on the \feii\ 1.64$\mu$m transition. 
    The middle panel shows the region that comprises the HH34 collimated jet and its driving source, 
    HH34 IRS. The top and bottom panels display the region around the two large
    bow shocks HH34 N and HH34 S, respectively, which are located at about 100\arcsec from HH34 IRS.
	Background areas display a mean rms of 9$\times$10$^{-19}$ erg s$^{-1}$ cm$^{-2}$ pixel$^{-1}$. The bottom contour mark the 
	5 $\sigma$ level in the upper and lower panels and the 10 $\sigma$ level in the central panel. The remaining nine 
	contours are traced using a logarithmic scale with the highest level corresponding to a flux 100 times greater than the first contour.
	We note that in spite of the large nebulosity associated with the source, the HH34 counter-jet is detected
    as a faint collimated emission.
       }

         \label{fig.2}
   \end{figure}

\section{Introduction}

The Herbig-Haro (HH) object HH34 and its related jet represent one of the most 
remarkable and well studied flow of matter from a young star. The HH34 object is
composed by a couple of large bow-shocks (HH34 S and HH34 N) that are symmetrically
displaced at a distance of about 0.2 pc with respect to the embedded 
driving source HH34 IRS. Associated with the two bow-shocks, a highly collimated
chain of knots originating from HH34 IRS forms a bright optical 
jet that points towards the HH34 S bow (B\"uhrke et al. 1988)

Imaging observations of a larger field 
around HH34 have revealed a number of additional bow shocks at larger 
distances, indicating that the total extent of this bipolar flow is almost 3 pc
(Eisl\"offel \& Mundt 1997; Devine et al. 1997).
The driving source HH34 IRS is a $\sim$0.5 \msun\ actively accreting protostar 
with \lbol$\sim$15 \lsun\ (Antoniucci et al. 2008).

The HH34 flow has been the subject of intense observational campaigns that
include spectro-photometry and kinematical measurements at both optical 
and IR wavelengths (e.g., Heathcote \& Reipurth 1992; Eisl\"offel \& Mundt 1992; Morse
et al. 1992, 1993; Reipurth et al. 2002; Beck et al. 2007). 
Detailed proper motion studies were conducted
with multi-epoch HST images, which were used to model the time-history of the flow
(Raga \& Noriega-Crespo 1998; Raga et al. 2012).
These studies show that the HH 34 jet is subject to
variations in velocity and jet axis direction, which might be due to precession of
the outflow induced by a possible companion.

Several IR spectroscopic investigations have been performed on the HH34 jet, showing
that it copiously emits in both \feii\ and H$_2$ transitions
(Podio et al. 2006, Takami et al.2006, Garcia Lopez et al. 2008, Davis et al. 2011).
The IR spectroscopy has revealed for the first time the presence of a 
counter-jet (Garcia Lopez et al. 2008), which has later been imaged with Spitzer (Raga et al. 2011).
The aim of this note is to complement the previous near-IR spectroscopic
investigations of HH34 with high quality imaging in \feii1.64$\mu$m, as 
obtained with the instrument LUCI (Seifert et al. 2003) mounted on the Large Binocular Telescope (LBT).
We compare the morphology of the jet in the IR with that of a commonly used optical
tracer like \sii. 
%
The \feii\ 1.64$\mu$m line traces dense ($\sim$ 10$^4$ cm$^{-3}$) shocked gas 
at low excitation ($T_{ex}$ = 11\,000 K): being in the infrared, it can
reveal details on the more embedded jet regions that remain obscured
at optical wavelengths. 

Previous imaging observations in \feii\ of HH34 were 
performed by Stapelfeldt et al. (1991) and Reipurth et al. (2000); this
latter observation was obtained with the HST-NICMOS camera. The sensitivity of these 
images was, however, not high enough to study the details of the IR jet
and identify the embedded counter-jet. Sensitive spectro-imaging in \feii\ were
obtained by Davis et al. (2011), but they cover only a small 
3\arcsec $\times$3\arcsec field around the HH34 IRS source.

In the present note, we describe the \feii\ morphology of the flow
in comparison with an optical \sii\ image taken with the HST. Such a comparison
has been improved by the application of a suitable deconvolution method (La Camera et al. 2014) 
to the seeing limited LBT image.
In addition, we measure the \feii/\sii\ ratio along the jet 
and relate it to iron depletion variations in the different shock episodes.


\section{Observations and reduction}

Imaging with LUCI ($4^{\prime}\times4^{\prime}$ field of view with a pixel scale of 0\farcs118/pix), which was
equipped with the \feii\ filter ($\lambda_{eff}=1.646$\um, $\Delta\lambda=0.018$\um), 
was carried out at the LBT on two different nights 
(Feb 19 and Apr 1 2013) using a dithering technique for a total combined exposure time of 30 minutes. 
The reduction of the two datasets was achieved using the LBT pipeline developed at the Rome Observatory,
in which the raw frames are first pre-reduced by subtracting a median stack
dark image and by applying a flat-field that is obtained by combining a set of sky flats (each subtracted by its own
dark and normalised by its own median background level). After this step, cosmetic masks are created to flag out 
saturated regions, cosmic rays events, and bad pixels; maps of background are then
subtracted from each image.
Finally, for each scientific frame, an accurate astrometric calibration is computed to correct for 
geometrical distortions; this allows the use of SWarp (Bertin et al. 2002) to both resample the 
processed images and create the final mosaic stack. 
Given that the seeing conditions were similar during the two nights of observations, the final \feii\ image was 
obtained from the combination of the two exposures. The Point Spread Function (PSF) 
full-width-at-half-maximum of this final image, which was measured on several field stars, 
is about 0\farcs85.


%
%
%

\section{Results and analysis}
\subsection{Large-scale morphology}
The central portion of the final image covering an area of 75\arcsec $\times$ 75\arcsec 
around HH 34 is shown in the middle panel of Fig.~1.
The collimated
southern jet (detected in optical images and composed by blue-shifted gas) emerges from 
the IR bright HH34 IRS source and is clearly visible. The northern counter-jet
is also detected, although at a much fainter level. 
The region around HH34 IRS is characterised by a large nebulosity, which
prevents a clear view of the red-shifted 
counter-jet close to the central source. An enhanced collimated emission is,
however, observed along the direction of the jet axis, which indicates that the
counter-jet extends close to the source, as already shown spectroscopically 
by Garcia-Lopez et al. (2008). 
The emission knots observed in the counter-jet are about an order of magnitude
fainter than the corresponding blue-shifted jet knots. Assuming this is due to the extinction,
this implies an A$_V$ of about 12-13 mag towards the brightest counter-jet
knots (at a distance of $\sim$ 15\arcsec from source). The extinction value in 
the jet blue-shifted and symmetrically displaced knots, as estimated through IR spectroscopy of
\feii\ line ratios, is about 1.5 mag (Podio et al. 2006). To account for the different extinctions,
a H$_2$ column density that is about a factor of ten higher in the counter jet than in the blue-shifted jet is therefore needed.
Such a higher column density in the red-shifted region is not supported
by millimeter maps of the region (Johnstone \& Bally, 2006). On the other
hand, the HH34 jet and counter-jet appear symmetric in the Spitzer IRAC images,
which are sensitive to H$_2$ emission (Raga et al. 2011). It is possible that the
counter-jet has a different excitation structure with the molecular component that is 
more enhanced with respect to the atomic one.

In addition to the HH34 jet, the 1.64$\mu$m image shows extended nebulosities that
originate from the IR sources IRS5 and Re22, but no collimated emission that is 
indicative of shock-excited gas is associated with the HH534 object
(Reipurth et al. 2002).
 
Symmetrically displaced at about 100\arcsec\, from the source, we detected both the 
 HH 34 N and HH34 S bow shocks (top and bottom panels of Fig.~1, respectively). They do not present 
 structures that are morphologically different from those observed in
\sii\ with HST. Figure 1 shows that the HH34 N bow shock appears to not be aligned with the
direction of the jet. This is consistent with the variations of the jet axis 
due to precession already observed in optical HST images.

\subsection{Image deconvolution and small scale morphology}

As we see from Fig.~1, the close surrounding of 
HH34 IRS is dominated by continuum emission from the
bright source and its reflection nebula. To reduce 
the contribution of the continuum emission and get 
information on the jet structure at resolutions comparable with
HST images, we have applied a deconvolution
algorithm optimised for separating point-like structures from
a diffuse extended continuum to the image, which is named \textit{multi-component Richardson-Lucy} (MC-RL) 
(La Camera et al. 2014).
The MC-RL method is based on the decomposition of the target as a sum of a 
point source (the star) and an extended source (the jet). By assuming Poisson noise, 
a regularisation term is added to the negative logarithm of the likelihood, 
which enforces smoothness of the jet component. A Richardson-Lucy like method is finally employed for 
the minimisation of the obtained function through iterations alternating between the two components.

   \begin{figure}[t]
   \centering
\includegraphics[angle=0,width=\hsize]{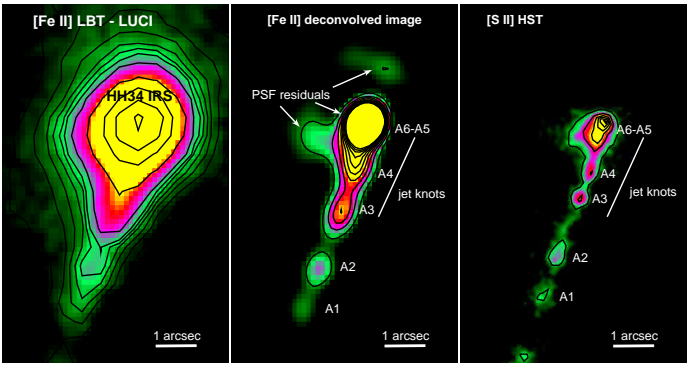}
      \caption{\feii1.64$\mu$m image of the central HH34 IRS region. The left and 
      middle panels show the original image and the image 
      deconvolved by adopting the algorithm described in La Camera et al. (2014), respectively. In the 
      left panel, a HST image in \sii\ at 6716+6730 \AA\, taken in 2007 is shown from
      comparison (Hartigan et al. 2011).}
         \label{fig.2}
   \end{figure}

The deconvolution algorithm was applied to the central region of the image,
which covers the entire HH34 collimated jet, by using 500 iterations. 
The PSF for the deconvolution was extracted from the average of three
(point-like) stars present in the image field by means of the Software Package
AIRY (Correia et al., 2002; La Camera et al., 2012). 

Figure 2 shows the inner part of the HH34 jet before and after the deconvolution. The PSF of the deconvolved image is around  
0\farcs24. For comparison, the figure also shows the most recent (2007) HST image in \sii\ at 6730\AA 
\,(Hartigan et al. 2011).
In the HST image the HH34 IRS continuum from the nebula is weak, and the 
jet is traced close to the source through several emission knots, 
which have been named after Reipurth et al. (2000) from A1 to A7, for decreasing distance from the
source. In the \feii\ deconvolved image, we are able to resolve knots from A1 to A5, which are
located at about 1\arcsec from the central source. The position of the 
\feii\ knot peaks are shifted with respect to the knots in the \sii\ image
due to the proper motion of the jet. Raga et al. (2012) have recently studied
the kinematics of the HH34 jet using HST images with a nine-year time
baseline. They found that the jet knots follow a ballistic motion and the
tangential velocity of the internal A knots is about 200 km\,s$^{-1}$.
Assuming a distance of 414 pc (Menten et al. 2007), we expect the knots
to have moved by about 0\farcs5 during the time-line between the \sii\ and \feii\
observations. In A1, A2, and A3, we measure a shift between 0\farcs4-0\farcs5,
which is consistent with previous proper motion results. This evidence implies in turn
that there is no large difference in the \feii\ and \sii\ morphologies,
at least at a half-arcsec resolution.

Figure 3 shows the \feii\ deconvolved image of the entire collimated jet compared to the \sii\ image.
Given the small number of field stars available for realignment, we estimate that the two images 
are registered with an accuracy of about 0\farcs2.
The jet morphology in the two tracers is again very similar. In the figure, the
main knot peaks detected in \feii\ are indicated with crosses and also
reported on the \sii\ image for comparison. All the knots visible in 
\sii\ are also detected in \feii\ with no appreciable variations in morphology
but with the expected shifts in position due to proper motion. 
These shifts range between 0\farcs4 and 0\farcs7 in the C-L knots with no
apparent trends with distance. Raga et al. (2012) derived that the tangential
velocity decreases from about 200 km\,s$^{-1}$ to about 170 km\,s$^{-1}$
going from the internal A knots to the more external C-E knots. We do not 
detect variations in the angular shifts corresponding to this velocity decrease,
which would correspond to about 0\farcs1. This is, however, lower than the estimated image alignment accuracy. 
   
   \begin{figure*}
         \centering
\includegraphics[angle=0,width=\hsize]{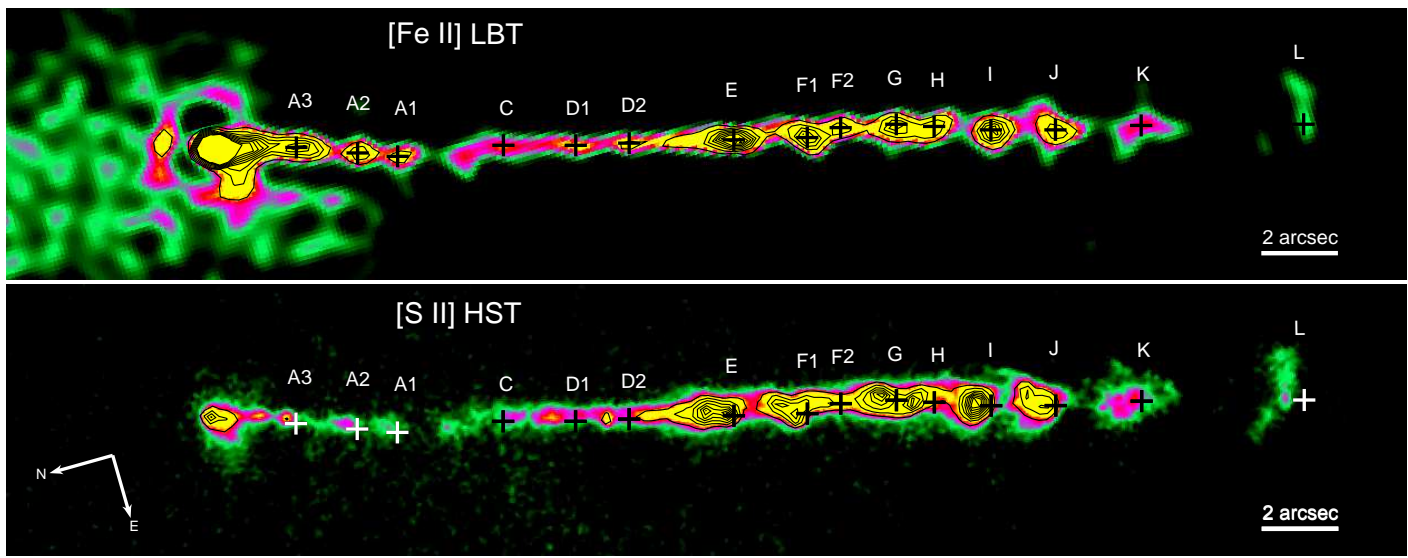}
      \caption{\feii1.64$\mu$m deconvolved image of the entire HH34 jet is shown
      in comparison with the HST image in \sii\ at 6716+6730 \AA. Both images are rotated
      for a better comparison. Crosses mark the peaks of the emission knots
      detected in the \feii\ image. The same positions are reported on the \sii\
      image to make evident the shifts due to the jet proper motion.}
         \label{fig.2}
   \end{figure*}
   
\subsection{\feii/\sii\ ratio and Fe abundance}

The LUCI narrow-band image has been flux calibrated by considering the 2MASS $K$-band
magnitude of four stars that are detected in the field. Zero point magnitude variations
of the order of 0.3 mag have been measured among the selected stars, which results
in a 20\% uncertainty on the absolute flux calibration of the image. 
Considering that the effective wavelengths of the Fe narrow-band and 2MASS $K$-band filters (1.646$\mu$m and 1.662$\mu$m, respectively) differ by less than 0.02$\mu$m and that the stars used for determining the zero point do not present peculiar spectral shapes in the near-IR range, we expect that the error related to the different bandwidths of the filters is negligible with respect to the local fluctuations of the sky background.
Table 1 reports the \feii\ fluxes measured on individual knots, by considering
an aperture with a radius of 0\farcs6 (representing the typical size of
the brightest knots) and centred on the knot peak (coordinates
are given in the Table). 

The \sii\ HST image has also been flux-calibrated to compare the \feii/\sii\
line ratios. With this aim, we used the star 2MASS J05353040-0627072,
which is the only optically bright and isolated star present in the field.
For this star, we adopted an $R$-band magnitude of 17.95$\pm$0.15 from the 
average of the $R$ photometry from the GSC 2.3.2 (Lasker et al., 2008)
and USNO-B1.0 (Monet et al., 2003) catalogues (18.1 and 17.8 mag, respectively). 
The relative uncertainty on flux calibration is therefore 15\%.
  
The \feii\ 1.64$\mu$m transition has a critical density of the order of 10$^{4}$
cm$^{-3}$ and an excitation temperature of $\sim$11\,000 cm$^{-1}$. In comparison, 
\sii\ (6716+6730 \AA) lines have critical densities of the order of 10$^3$ cm$^{-3}$
and excitation temperatures of the order of 20\,000 cm$^{-1}$. Therefore, \feii\ is more
sensitive than \sii\ to dense gas at low excitation. Nisini et al. (2005)
compared the relative intensity profiles of \feii\ and \sii\ lines expected 
in the post-shocked gas, showing that the emission zone of the \feii1.64$\mu$m line 
overlaps with that of the \sii\ optical lines, although it is broader and 
covers a larger area in the post-shocked region. However, these spatial differences are well
below the resolution of our observations, being the cooling zone of the post-shocked
gas only about 10$^{14}$ cm wide (i.e. 0\farcs02 at 414 pc) (Hartigan et al. 1994).

Given the above considerations, the \feii1.64$\mu$m/\sii\ (6716+6730 \AA)
line ratio is sensitive to the shock physical conditions, in
particular for temperatures up to 20\,000 K and in the density range between
10$^3$ and 10$^5$ cm$^{-3}$ (see Fig. 4). This ratio is, however, also sensitive
to the relative Fe/S abundance in the shocked gas. Iron, being a refractory element,
is depleted on grains in the dusty interstellar medium. If jets are dusty, shocks can erode the grains
and release Fe in gaseous form in a percentage that depends on the shock strength
and pre-shock density (e.g., Draine 2003). Several studies have shown that the 
abundance of Fe gas along jets is indeed much lower than solar with percentages
of depletion that considerably change among shocks and range from $\sim$ 10\% up to
90\%  (i.e. Nisini et al. 2002, 2005, Podio et al. 2006, Giannini et al. 2008, 2013).
In particular, Podio et al. (2006) derived that the abundance of Fe gas along the HH34 jet, which 
is measured using the ratio of \feii1.64$\mu$m/\pii1.18$\mu$m lines, is
only about 10\% of the solar value. In Fig.~4, we compare the \feii1.64$\mu$m/\sii\ (6716+6730 \AA)
ratio in the various knots along the jet with the predicted values by assuming solar Fe/S abundance ratio.
Abundance values were taken from Asplund et al. (2006). For this comparison, we have 
measured the \sii\ intensity in the knots by considering the peak shifts with respect to
the positions given in Table 1 due to the proper motion.
\begin{table}
\caption{\feii\ 1.64$\mu$m line fluxes.}             
\label{table:1}      
\begin{tabular}{l c c c}        
\hline\hline                 
knot & R.A. (J2000) & DEC (J2000) &  $F^a$ \\    
     & ~~$h$~~~$m$~~$s$ &~~~~\degr~~~~\arcmin~~~~\arcsec & (10$^{-15}$ erg\,s$^{-1}$\,cm$^{-2}$)\\
\hline                        
A3 & 5:35:29.89 & -6:27:00.6 & 6.1 \\
A2 & 5:35:29.92 & -6:27:02.0 & 2.7 \\
A1 & 5:35:29.95 & -6:27:03.2 & 1.5 \\
C & 5:35:29.99 & -6:27:05.8 &  1.3 \\
D1 & 5:35:30.02 & -6:27:07.5 & 1.3 \\
D2 & 5:35:30.04 & -6:27:09.0 & 1.4 \\
E & 5:35:30.09 & -6:27:11.6 &  8.8 \\
F & 5:35:30.11 & -6:27:13.5 &  5.9 \\
G & 5:35:30.14 & -6:27:15.7 & 5.0 \\
H & 5:35:30.16 & -6:27:16.8 & 3.8 \\
I & 5:35:30.19 & -6:27:18.3 & 5.7 \\
J & 5:35:30.22 & -6:27:19.9 & 3.4  \\
K & 5:35:30.26 & -6:27:22.1 & 1.3 \\
L & 5:35:30.3 & -6:27:26.3 &  0.4 \\
\hline                                   
\end{tabular}
\\
$^a$Fluxes measured in a 0\farcs6 radius beam centred at the given coordinates.
Flux calibration uncertainty is of the order of 20\%.
\end{table}
Observed values have also been corrected for extinction by assuming the A$_V$ values estimated
by Podio et al. (2006), which are namely A$_V$=7 mag in the A knots and A$_V$=1.5 mag in the 
more external C-L knots (see also Garcia Lopez et al. 2008). 
An error of about 1 mag can be considered for these visual extinction estimates: this is basically due to the
uncertainty on the rate coefficients of the [Fe II] lines used for the A$_V$ determination
(see, e.g., Giannini  et al. 2008).
The figure shows that the observed
ratios are always significantly lower than the predicted values, irrespective of the
assumed physical conditions. This agrees with Fe being largely depleted from gas phase.
We point out that the presented comparison assumes that \feii\ and \sii\ come
from the same spatial region. However, as discussed above, \feii\ covers
a larger post-shocked region with respect to \sii, so the observed values
should be corrected for their relative filling factors before comparison. This correction
would make the observed ratios even lower, thus exacerbating the difference with
models that assume solar values.

      \begin{figure}
      \vspace{-1.0cm}
\includegraphics[width=\hsize]{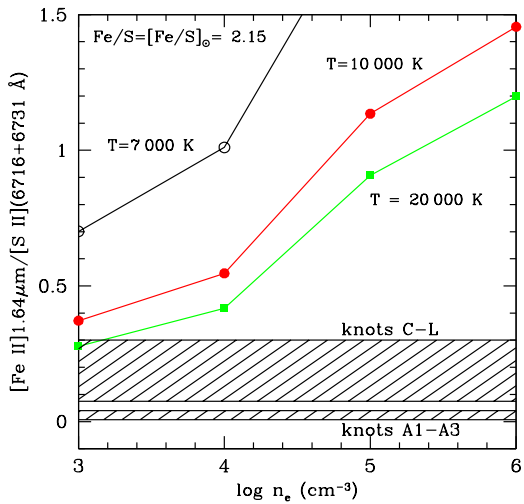}
      \vspace{-0.5cm}
      \caption{\feii1.64$\mu$m/\sii(6716+6731\AA) intensity ratio, as predicted by
      NLTE statistical equilibrium calculations, as a function of the electron density
      and for temperatures of 10\,000 K and 20\,000 K. Solar abundances from Asplund et al. (2006) 
      are assumed. The
      dashed areas indicate the range of extinction corrected values measured
      in the internal knots A and the external knots from C to L. 
      The reported ranges also include the uncertainties from both image calibrations and extinction.      
      The measured ratios are always below the predicted value, irrespective to the assumed temperature 
      and density, which indicates a Fe abundance lower than solar.}
         \label{fig.2}
   \end{figure}

%
%
%

From the ratios of several optical/IR forbidden lines, Podio et al. (2006) and Garcia Lopez et al. (2008) 
determined that the density in the HH34 jet varies from $\sim$10$^{4}$
to $\sim$1-3\,10$^{3}$ cm$^{-3}$ in the inner A and outer C-L knots, respectively, with a temperature
in the range 6000-14\,000 K. Assuming these conditions, from Fig.~4 we infer that
the Fe gas phase abundance is $\sim$1-10\% Fe$_\odot$ for knot A and 
$\sim$15-90\% Fe$_\odot$ for knots C-L. These results show evidence that a larger Fe depletion
is found in the inner knots with respect to the rest of the jet. This dependence
of Fe depletion on the distance from the driving source was also observed in the HH1 jet 
(Nisini et al. 2005). Other refractory species, like Ca, show the same trend of
decreasing gas-phase abundance in the inner jet regions (Podio et al. 2009).
These trends might be due to a higher density or lower shock velocity in the inner regions,
which would diminish the efficiency of dust disruption.


      

%


\begin{acknowledgements}
    We kindly thank Pat Hartigan for providing the \sii\ HST image.\\
    We acknowledge the support from the LBT-Italian Coordination Facility for the execution of observations, data distribution and reduction.\\
    The LBT is an international collaboration among institutes in the United States, 
    Italy, and Germany. LBT corporation partners are: the University of Arizona on behalf of the Arizona university system;
    the Istituto Nazionale di Astrofisica, Italy; the LBT Beteiligungsgesellschaft, Germany, representing the Max-Planck Society,
    the Astrophysical Institute of Potsdam, and Heidelberg University; The Ohio State University, and the Research Corporation, on
    behalf of the University of Notre Dame, University of Minnesota, and University of Virginia.\\
    The Guide Star Catalogue-II is a joint project of the Space Telescope
    Science Institute and the Osservatorio Astronomico di Torino.\\
    The USNO-B1.0 catalog was created by Dave Monet and collaborators at http://www.nofs.navy.mil/data/fchpix/
\end{acknowledgements}


\end{document}